\documentclass[aps,preprint,amsmath,amssymb]{revtex4}
\usepackage{graphicx,color}

\def\be{\begin{eqnarray}}
\def\ed{\end{eqnarray}}
\def\non{\nonumber}

\begin{document}

\title
{\Large \bf Resolution to neutrino masses, baryon asymmetry in
leptogenesis and cosmic-ray anomalies }

\author{ \bf  Chuan-Hung Chen$^{1,2}$\footnote{Email:
physchen@mail.ncku.edu.tw}}

\affiliation{ $^{1}$Department of Physics, National Cheng-Kung
University, Tainan 701, Taiwan \\
$^{2}$National Center for Theoretical Sciences, Hsinchu 300, Taiwan
 }

\date{\today}

\begin{abstract}
By extending the lepton sector of standard model to include one
sterile neutrino and two sets of new Higgs doublets and right-handed
neutrinos, denoted by $(\eta_1, N_1, N_3)$ and $(\eta_2, N_2, N_4)$,
with two $Z_2$ symmetries, the puzzles of neutrino masses,
matter-antimatter asymmetry and cosmic-ray excess observed by
Fermi-LAT and PAMELA can be resolved simultaneously. The characters
of the model are: (a) neutrino masses arise from type-I and
radiative seesaw mechanisms; (b) leptogenesis leading to baryon
asymmetry at the energy scale of $O(1- 10\rm TeV)$ could be realized
through soft $Z_2$ symmetry breaking effects; and (c) the conditions
of small couplings for a long-lived dark matter could be achieved
naturally through loop corrections due to the same soft symmetry
breaking effects. The candidate for dark matter in leptophilic
decays could fit the Fermi-LAT and PAMELA data well.

\end{abstract}

\maketitle

One of the strongest motivations for
new physics
is due to
the unsolvable problems in the standard model (SM), such as the
origin of neutrino masses, matter-antimatter asymmetry and dark
matter,
which are the phenomena
observed well,
but less understood.
Recently, three  astonishing experiments
PAMELA~\cite{PAMELA}, ATIC~\cite{ATIC} and Fermi-LAT
\cite{Fermi-LAT} in cosmic-ray measurements reheated the issue of the
unknown stuff, in which the first one indicates anomaly in positron
flux ratio while the later two observe excess in electron+positron
($e^+ e^-$) flux. Although Fermi-LAT's results don't display the
bump as shown at ATIC in the $e^-e^+$ flux,  a clear excess
at Fermi with high precision can not be denied. Promptly, it is
proposed that one of possible resources for generating the exotic
events in the flux of electrons and positrons is ascribed to the
dark matter annihilations and/or decays \cite{DManni,DMdecay,CGZ}.

PAMELA observes not only on the positron flux ratio but also on the
antiproton flux ratio, however, the resulting evidences show no
significant excess on the latter. Accordingly, if the excess
observed at Fermi and PAMELA stems from the dark matter, the dark
stuff should presumably be leptophilic. Furthermore, if
the
matter-antimatter asymmetry arises from the so-called
leptogenesis where the baryon asymmetry originates from the lepton
asymmetry \cite{FY}, the issues of neutrino masses, baryon asymmetry
and Fermi/PAMELA puzzles intriguingly are all related to the lepton
sector. Based on this inference, it is interesting to investigate
how these unsolved problems can be explained within a simple
model uniformly.

Although dark matter annihilations could provide the source for the
Fermi/PAMELA anomalies, it is inevitable that an enhanced
boost factor of a few orders of magnitude, such as Sommerfeld
enhancement \cite{BoostFactor}, near-threshold resonances and
dark-onium formation \cite{Dark-onium}, has to be introduced.
To evade of
introducing
unnecessary effects,
we
will focus on the mechanism of dark matter decays. Moreover,
to get a long-lived dark matter, say $O(10^{26} s)$, usually we
have to fine-tune either the new couplings to be tiny or the new
scale in intermediated state to be as large as the scale of grand unified
theories (GUTs). Hence, in order to fulfill our purpose, the
subject we face is not only how the model provides a new CP
violating (CPV) mechanism in the lepton sector so that lepton number
asymmetry could be converted to the baryon asymmetry by the
nonperturbative sphalerons \cite{KRS}, but also how a long-lived
dark matter exists naturally while solving Fermi/PAMELA anomalies.

To solve the mentioned problems, we extend the SM by including
two extra Higgs doublets $\eta^T_i=(\eta^0_i, \eta^-_i)$ (i=1,2)
with hypercharge $Y=-1$ and five right-handed neutrinos denoted by
$N_{1-4}$ and $N$. Besides, we introduce two $Z_2$ discrete
symmetries so that $\eta_1$ and $N_{1,3}$ are transformed by
 \be
 \eta_1\to -\eta_1,\ \ \ N_{1,3}\to -N_{1,3}\,,
 \ed
while $\eta_2$ and $N_{2,4}$ follow
 \be
 \eta_2\to -\eta_2\,, \ \ \ N_{2,4}\to -N_{2,4}
  \ed
under the first and second $Z_2$ symmetry, respectively. The
unmentioned fields in above equations denote invariance in each
transformation. In addition, $\eta_{1,2}$
do not have vacuum
expectation values (VEVs) when the discrete symmetries are exact.
Accordingly, the Yukawa sector could be found as
 \be
-{\cal L}_{Y} &=&\bar L_i Y^{E}_{ij}  \tilde{H}  \ell_{jR}
+ \bar L_j  Y^N_j  H N  +\bar L_i y_{i\alpha} \eta_1 N_\alpha \non \\
&&+ \bar L_i y_{i\beta} \eta_2 N_\beta + \frac{m_N}{2} N^T C N\non \\
&&  + \sum^{4}_{k=1} \frac{m_{N_k}}{2} N^T_k C N_k + H.c.\,,
\label{eq:yukawa}
 \ed
where $L^T =(\nu_\ell, \ell)_L$ denotes the $SU(2)_L$ doublet
leptons, $H^T=(H^0, H^-)$ is the SM Higgs doublet, $Y^E_{ij}$,
$Y^{N}_j$ and $y_{i\alpha(\beta)}$ with $i,j=1-3$ and
$\alpha=1,3\, (2,4)$ stand for the Yukawa couplings. For
simplicity, we have chosen $N_i$ as the diagonalized states
before the electroweak symmetry breaking (EWSB). Since the right-handed
neutrino $N$ directly couples to the SM Higgs and ordinary leptons, the
masses of neutrinos could be induced after the EWSB. Therefore, $N$
could lead to non-zero neutrino masses by the type-I seesaw mechanism
\cite{SeaSaw}, i.e., we have to set $m_{N}\gg m_{N_{i}}$. Moreover,
because of interactions associated with charged leptons being
$\bar\ell_L N_i \eta^+_i$, in order to explain the Fermi/PAMELA
data, the decaying dark matter should be the fermionic particle.
Here, we take
$N_{2}$ as the candidate. Thus, the mass relations
are required to obey $ m_{\eta_{1(2)}}> m_{N_{1,3(2)}}$ and
$m_{N_2}>m_{N_{1}}$. We will see clearly later that if $m_{N_4}\gg
m_{N_3}>m_{\eta_2}$, the role of $N_{3}$ is responsible to the
leptonic CP asymmetry (CPA) in leptogenesis.

Since $N_{2}$ is protected by  the $Z_2$-symmetry and the decay is forbidden by
kinematics, so far it is still a stable particle. We note that $N_3$
and $N_4$ are allowed kinematically to decay
through the channels $N_3\to
N_2 \ell^+ \ell^-$ and $N_4\to \ell \eta_2$, respectively. In order
to make the dark matter unstable, we ascribe that the origin of
the unstable dark matter is from the $Z_2$ soft breaking terms, given by
 \be
-{\cal L}_{\rm soft}=\mu_{34} N^T_3 C N_4 + H.c \,. \label{eq:soft}
 \ed
Due to the soft breaking effects being associated with heavier
right-handed neutrinos, it is clear that although $Z_2$ symmetries
have been broken, the breaking effects will not open a sizable
decaying channel for $N_2$. Hence, the unstable dark matter
could be a long-lived one.

\begin{figure}[htbp]
\includegraphics*[width=2. in]{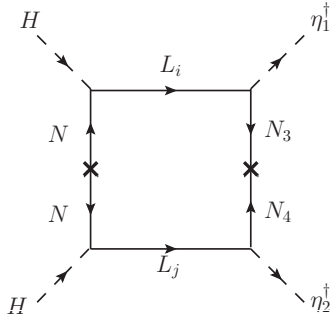}
\caption{Box diagram for dimension-4 term $\left(\eta_1^\dagger H
 \right)
 \left( \eta^\dagger_2 H
 \right)$.}
  \label{fig:box}
\end{figure}

We now start to discuss in turn how the puzzles of small neutrino
masses, baryon asymmetry and
Fermi-LAT/PAMELA are solved in this model. \\
{\it \bf Neutrino Masses}: By the first term of
Eq.~(\ref{eq:yukawa}), the neutrino mass matrix could be expressed
by
 \be
\left(m_\nu\right)^{\rm Type-I}_{ij}\approx  -m_{Di} m^{-1}_N
m^{\dagger}_{Dj}, \ \  \ m_{Dj}=\frac{v}{\sqrt{2}} Y^{N}_{j}\,,
\label{eq:typeI}
 \ed
where we have made the expansion in terms of $m_D/m_N$ and used
$\langle H \rangle=v/\sqrt{2}$ as the vacuum expectation value (VEV)
of the Higgs field. To explain the neutrino masses, with
$Y^{N}_{i(j)}\sim O(1)$ we see that $m_N\sim 3\times 10^{15}{\rm
GeV}/(m_{\nu}/1 {\rm eV})$ as expected by the conventional seesaw model.
Beside the type-I seesaw mechanism, the masses of neutrinos could be
also induced by radiative corrections in which the corresponding
Feynman diagram is illustrated in Fig.~\ref{fig:box}(a). With the
relevant quartic term in the Higgs potential, written by $\lambda_5/2
\left(H^\dagger \eta_{1,2} \right)^2$, the mass matrix through
one-loop is given by \cite{MA}
  \be
\left(m_\nu \right)^{\rm rad}_{ij}&\approx & \frac{\lambda_5
v^2}{8\pi^2} \left( \frac{y_{i1}
y_{j1}m_{N_1}}{m^2_{\eta_1}}\right.\non\\
  &+&\left.  \frac{y_{i3}
y_{j3}m_{N3}}{m^2_{\eta_1}} + \frac{y_{i2}
y_{j2}m_{N2}}{m^2_{\eta_2}}\right)\,, \label{eq:rad}
  \ed
where we have adopted $m_{\eta_{1(2)}}>m_{N_{1,3(2)}}$. Because
$N_{4}$ is much heavier than other particles, we have
ignored its contributions. With $m_{\eta_1}\sim 8$ TeV,
$m_{\eta_2}\sim m_{N_3}\sim 4$ TeV, $m_{N_{1}}\sim m_{N_{2}}\sim O(1
{\rm TeV})$, $\lambda_5\sim O(10^{-4})$ and $y_{i3}\sim O(10^{-2})$,
the neutrino mass by the dominant effects could be $O(1{\rm eV})$.
Intriguingly, if we combine the type-I and radiative seesaw
mechanisms together, it is found that some cancelations could occur
between both. As a result, the constraint on the couplings of
$\lambda_5$ and $y_{i1(2,3)}$ could be somewhat relaxed. Besides,
the $m_{\nu}\sim 1$eV resulted in each seesaw mechanism can be
reduced to $O(0.01-0.1\rm eV)$ by the cancelations.

\begin{figure}[htbp]
\includegraphics*[width=3.5 in]{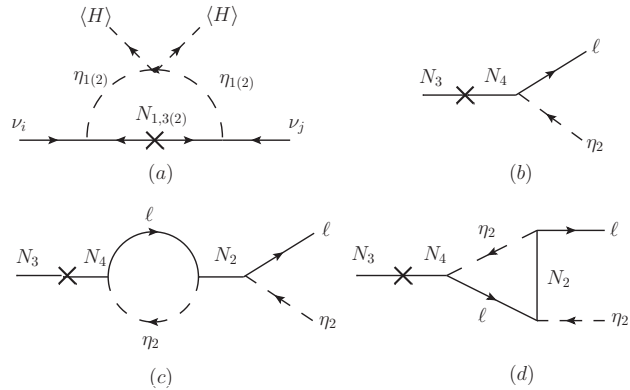}
\caption{Feynman diagrams of (a) Neutrino masses by radiative corrections and  (b), (c) and
(d)  the  $N_3\to \ell \eta_2$ decay at tree and loop levels.
} \label{fig:lepto}
\end{figure}

{\it \bf Leptogenesis}: Since $N_3$ plays the role of the lepton
asymmetry at the energy scale of $O(1-10\rm TeV)$, to satisfy the
out-of-equilibrium condition, the decay rate of $N_3$ should be less than the
Hubble constant H at the temperature of $m_{N_3}$, given by
\cite{Sakharov}
 \be
 \Gamma_{N_3}< H(T=m_{N_3})=\sqrt{\frac{4\pi^3 g_*}{45}} \frac{T^2}{M_{\rm
 Planck}}|_{T=m_{N_3}}
 \ed
where $g_*\approx 100$ denotes the number of active degrees of
freedom and $M_{\rm Planck}\sim 10^{19}$ GeV is the Planck scale. As
mentioned early, $N_3$ could decay via the three-body channel of
$N_3\to N_1 \ell^{+} \ell^-$. We check that by the phase space
suppression and with $y_{i3} y_{i1}\sim 10^{-5}$, the rate for the
three-body decay is around a factor of 5 smaller than Hubble
constant. In addition, by the soft breaking effects of
Eq.~(\ref{eq:soft}),  $N_3$ will decay to $\ell \eta_1$ through the
mixing with $N_4$, sketched by Fig.~\ref{fig:lepto}(b). The rate
will depend on the parameters of $\mu_{34}$, $m_{N_4}$ and $y_{i4}$.
By calculating the two-body decay rate of Fig.~\ref{fig:lepto}(b)
and with $m_{N_3}=4$ TeV, the constraint on the parameters is
 \be
\left|\frac{\mu_{34}}{m_{N_4}}y_{i4}\right|<5\times
10^{-7}\left(\frac{m_{N_3}}{4\rm TeV}\right)\,. \label{eq:mu34}
 \ed
By employing small $\mu_{34}/m_{N_4}$, the decay rate less than
the
Hubble constant for $N_3$ of $O(\rm TeV)$ can be naturally
accomplished.

\begin{figure}[htbp]
\includegraphics*[width=2.5 in]{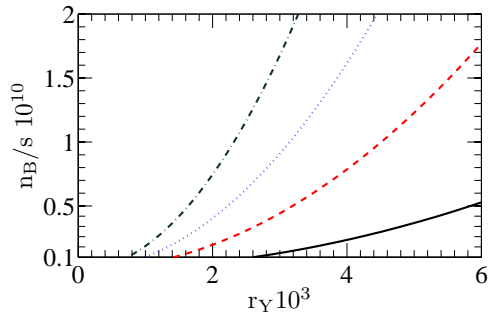}
\caption{Baryon asymmetry by leptogenesis,
where the solid, dashed, dotted and dash-dotted lines represent
$N_2=2$, 2.5, 3.0 and 3.5 TeV, respectively, with $N_3=4$ TeV.}
\label{fig:ba}
\end{figure}

Another necessary condition to achieve baryon asymmetry is the CPA,
defined by
 \be
A_{CP}&=& \frac{\Gamma(N_3\to \ell \eta^\dagger_2) -\Gamma(N_3\to
\ell^c \eta_2)}{\Gamma(N_3\to \ell \eta^\dagger_2) +\Gamma(N_3\to
\ell^c \eta_2)}\,.
 \ed
A nonzero direct CPA should involve CPV and CP conserving (CPC)
phases simultaneously. Here, the complex Yukawa couplings
$y_{i\alpha}$ can provide the new CPV source.
Therefore, we need a
new mechanism to generate the physical CPC phase. It has been well
known that one-loop effects could produce the CPC phase when the
on-shell condition of particle in the internal loop is satisfied. In
the model the CPC phases could be induced via self-energy \cite{FPS}
and vertex corrections \cite{FY} illustrated in
Fig.~\ref{fig:lepto}(c) and (d).
Therefore, the CPAs via self-energy (S) and vertex (V) diagrams are
respectively found to be~\cite{Hambye}
 \be
A^{S}_{CP}(N_3) &\approx & -\frac{1}{8\pi} r^2_Y \sin\delta
\frac{\sqrt{x_2}}{x_2-1}\,,\non\\
A^{V}_{CP}(N_3) &\approx&-\frac{1}{8\pi}
r^2_Y \sin\delta \non\\
&\times& \sqrt{x_2}\left[(1+x_2)\log(1+1/x_2)-1 \right]
  \ed
with $x_2=(m_{N_2}/m_{N_3})^2$ and $r^2_Y
\sin\delta=Im(y^*_{i4}y_{i2})^2/|y_{i4}|^2$. $\delta$ is taken as
the CPV phase and the repeating index $i$ stands for the summation
in lepton flavors. Following the relation of the observed baryon
asymmetry and the lepton asymmetry, formulated by \cite{KS-HT}
 \be
\left(\frac{n_B}{s}\right)_{\rm obs}= \frac{28}{79}
\frac{n_B-n_L}{s}\,,\ \
\frac{n_L}{s}\simeq \frac{ n_{\gamma}}{2s}A_{CP}
 \ed
where $s$ and  $n_{B(L,\gamma)}$ are the entropy and baryon
(lepton,$\gamma$) density of the universe, respectively,
we obtain
 \be
\left(\frac{n_B}{s}\right)_{\rm obs}\approx -\frac{1}{12g_*}
[A^{S}_{CP}(N_3)+A^{V}_{CP}(N_3)]\,,
 \ed
 where
 we have used $s=g_*T^3(2\pi^2/45)$ and
$n_\gamma=2T^3/\pi^2$. With $|\sin\delta|=0.3$, the baryon asymmetry
as a function of $r_Y$ is presented in Fig.~\ref{fig:ba}, where the
solid, dashed, dotted and dash-dotted lines represent $N_2=2$, 2.5,
3.0 and 3.5 TeV, respectively, with $N_3=4$ TeV.
According to the results, $r_{Y}\sim |y_{i2}|\sim O(10^{-3})$ not
only satisfies the criterion for radiative neutrino masses but also
explains the baryon asymmetry in leptogenesis.

{\it \bf Fermi-LAT/PAMELA}: Due to $N_{3,4}$ being heavier than
other particles, the soft breaking interactions introduced in
Eq.~(\ref{eq:soft}) cannot directly lead to $N_2$ decays. However,
by combining the soft breaking effects with Yukawa couplings, we
find that a finite dimension-4 hard operator for the mixture of
$\eta_1$ and $\eta_2$ could be induced at one-loop level, sketched
in Fig.~\ref{fig:box}. The resultant contributions to the scalar
potential is given by
 \be
 \delta V_{\rm hard}&=& C_{h}  \left(\eta^\dagger_1 H  \right)
 \left( \eta^\dagger_2  H
 \right)+ H.c\,,
 \ed
with
 \be
%
C_{h} &=&\frac{1 }{16\pi^2} \left( y^*_{i3}Y^{N}_{i} \right)\left(
 y^{*}_{j4}Y^{N}_{j}\right)\frac{\mu_{34}}{ m_N}\,,
 \ed
where we have neglected the terms associated with $m_{N_{3,4}}/m_N$
and used the Einstein summation convention for the indices $i$ and $j$.
If we simply ignore the summation in indices $i$ and $j$, by
combining with Eq.~(\ref{eq:typeI}),  the mixing effect of $\eta_1$
and $\eta_2$, denoted by $\mu^2_{12}$, could be simplified to be
 \be
\mu^2_{12} \sim \frac{1}{16\pi^2} (m_{\nu})_{ij}y^*_{i3} y^*_{j4}
\mu_{34}\,. \label{eq:mu12}
 \ed
With  using $y_{i3}\sim 10^{-2}$ and
$m_{N_4}=5\times 10^{4}$ TeV and
 following Eqs.~(\ref{eq:mu34}) and (\ref{eq:mu12}), we
immediately have $\mu^2_{12}< 1.5\times 10^{-12}\rm GeV^{2}$. Since
$Z_2$ symmetries will be restored while $\mu_{34}$ vanishes, the
smallness of $\mu^2_{12}$ is still nature in technique.
Consequently, the lifetime for $N_2$ dictated by $N_2\to N_1
\bar\ell \ell$ can be found by
 \be
\tau_{N_2}\simeq \frac{ 2^{9}\pi^3}{3|y_{i2}y^*_{j1}|^2}\frac{\left(
m^2_{\eta_1} m^2_{\eta_2}/\mu^2_{12}\right)^2}{m^5_{N_2}
\left(1-m^2_{N_1}/m^2_{N_2} \right)^4}\,.
 \ed
Adopting $m_{\eta_{1(2)}}=8(4)$ TeV, $m_{N_{1(2)}}=0.5(2)$ TeV,
$y_{i1}y_{i2}\sim 7.5\times 10^{-6}$ and the obtained upper value of
$\mu^2_{12}$, we get $\tau_{N_2}\approx 2.5\times 10^{26}$ s.
Clearly, the values of the parameters in the model could be
compatible with the phenomena of neutrino masses, leptogenesis and
long-lived dark matter.

Since the long-lived dark matter could only decay in leptophilic,
for illustration, we simply consider the electron-positron pair and
one electron (positron) and antimuon (muon) as the possible final
states, in which antimuon (muon) will further decay to positron
(electron) by SM weak interactions. Due to three possible channels
involved, where the associated parameters are proportional to
$|y_{e2}y^*_{e1}|^2$, $|y_{e2}y^*_{\mu1}|^2$ and
$|y_{\mu2}y^*_{e1}|^2$, the energy spectrum for electron (positron)
can be formulated by
 \be
 \frac{dN_{e^\pm}}{dE}&=&\frac{1}{N}\left(c^2_\alpha\frac{d\Gamma_{ee}}{dE}+
 c^2_\beta \frac{d\Gamma_{e\mu}}{dE} +
 c^2_\gamma\frac{d\Gamma_{\mu e}}{dE}\right)\,,
 \ed
 with
 \be
 \frac{d\Gamma_{e e}}{dE}  &=&  E^2\left(
\frac{3}{2}\Delta m^2_{21} -\frac{8}{3} m_{N_2}E \right)\,,\non\\
\frac{d\Gamma_{e\mu}}{dE} &=& F_{e\mu}(E) + \int^{m_{N_2}/2}_{m_\mu}
dE_\mu F_{\mu e}(E_\mu) \frac{d|M_\mu|^2(E_\mu, E)}{dE}\,,
\non \\
%
\frac{d\Gamma_{\mu e}}{dE} &=& F_{\mu e}(E)
+\int^{m_{N_2}/2}_{m_\mu} dE_\mu F_{e\mu}(E_\mu)
\frac{d|M_\mu|^2(E_\mu, E)}{dE} \non
 \ed
where $N$ is the normalization, $c^2_\alpha+c^2_\beta+c^2_\gamma=1$,
$\Delta m^2_{21}=m^2_{N_2}-m^2_{N_1}$, $F_{e\mu}(E)=E^2( \Delta
m^2_{21} -2 m_{N_2}E )$, $F_{\mu e}(E)=E^2( \Delta m^2_{21}/2 -2/3
m_{N_2}E)$ and
 \be
\frac{d|M_\mu|^2(E_\mu, E)}{dE} &=& \frac{ G^2_F}{6\pi^3}
\frac{1}{|\vec p_\mu|} \left\{ \frac{4}{3}\left[E^3\left( E_\mu
-|\vec p_\mu|\right)^3 -\left( \frac{m^2_\mu}{2}\right)^3
\right]\right. \non\\
&&\left. -\frac{3}{2}m^2_{\mu} \left[ E^2 \left(E_\mu -|\vec p_\mu|
\right)^2 -\left( \frac{m^2_\mu}{2}\right)^2 \right]\right\}\,.
 \ed
For $d\Gamma_{e\mu}/dE$ and $d\Gamma_{\mu e}/dE$, the allowed energy
range for muon and electron are found to be $m_\mu\leq E_\mu \leq
m_\chi/2$ and $0\leq E \leq m^2_\mu/2(E_\mu - |\vec p_\mu|)$,
respectively. Since the experiments measure the flux of cosmic-rays,
the used formalism to estimate the flux from the new source is given
by
   \be
\Phi^{N_2}_{e^\pm}= \frac{c}{4\pi}\frac{1}{ m_{N_2} \tau_{N_2}
}\int\limits_0^{m_{N_2}/2}dE^\prime
G(E,E^\prime)\frac{dN_{e^\pm}}{dE^\prime}
 \ed
with $c$ being the speed of light. For numerical estimations, we
adopt the result parametrized by \cite{IT_JCAP07}
 \be
G(E,E^\prime) \simeq
\frac{10^{16}}{E^2}\exp[a+b(E^{\delta-1}-E^{\prime\delta-1})]\theta(E^\prime-E)
\quad [{\rm cm}^{-3}{\rm
    s}]
 \ed
with $a=-0.9809$, $b=-1.1456$ and $\delta=0.46$.

\begin{figure}[hptb]
\includegraphics*[width=5 in]{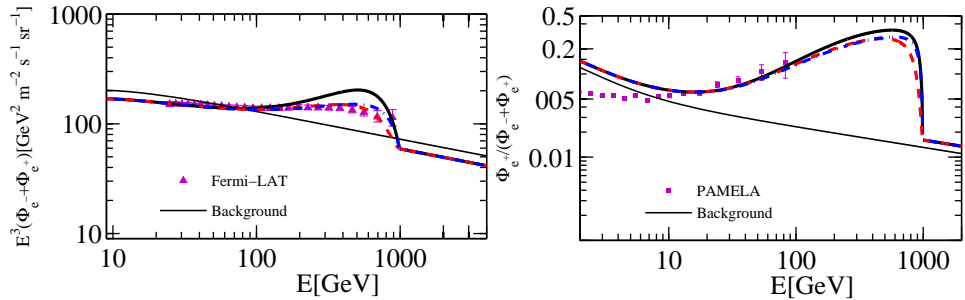}
\caption{electron+positron  flux (left) and the ratio of positron
flux to electron+positron flux (right), where the thick solid,
dashed and dash-dotted lines represent  $(c^2_\alpha,
c^2_\beta, c^2_\gamma)=(1,0,0)$, $(0,1,0)$ and $(0,0,1)$, respectively. }
  \label{fig:FP}
\end{figure}

For including the primary and secondary electrons and secondary
positrons, we use the parametrizations, given by
\cite{BE,Moskalenko}
 \be
    \Phi_{e^-}^{\rm prim}(E)&=&\frac{0.16E^{-1.1}}{1+11E^{0.9}+3.2E^{2.15}} \quad
    [{\rm GeV}^{-1}{\rm cm}^{-2}{\rm s}^{-1}{\rm sr}^{-1}], \non \\
    \Phi_{e^-}^{\rm sec}(E)&=&\frac{0.7E^{0.7}}{1+110E^{1.5}+600E^{2.9}+580E^{4.2}} \quad
    [{\rm GeV}^{-1}{\rm cm}^{-2}{\rm s}^{-1}{\rm sr}^{-1}], \non \\
    \Phi_{e^+}^{\rm sec}(E)&=&\frac{4.5E^{0.7}}{1+650E^{2.3}+1500E^{4.2}} \quad
    [{\rm GeV}^{-1}{\rm cm}^{-2}{\rm s}^{-1}{\rm sr}^{-1}]
    \label{eq:bg}
 \ed
where $\Phi^{\rm prim (sec)}$ denotes the primary (secondary) cosmic
ray. Accordingly, the total electron and positron fluxes are defined
by
 \be
\Phi_{e^-}&=&\kappa \Phi^{\rm prim}_{e^-}+\Phi^{\rm
sec}_{e^-}+\Phi^{N_2}_{e^-}\,, \non \\
\Phi_{e^+}&=&\Phi^{\rm sec}_{e^+}+\Phi^{N_2}_{e^+}\,.
\label{eq:flux_tot}
 \ed
Here, according to Refs.~\cite{Moskalenko} and \cite{BEFG}, we have
regarded the normalization of the primary electron flux to be
undetermined and parametrized by  the parameter of $\kappa$. The
value of $\kappa$ is chosen to fit the data. Before introducing the
source of the primary positron,  $\kappa$ is set to be 0.8. Taking
$m_{N_{2(1)}}=2(0.2)$ TeV and $\kappa=0.65$, we present the $e^-
e^+$ flux and ratio of fluxes $e^+/(e^- + e^+)$ by $N_2$ decays in
Fig.~\ref{fig:FP}, in which the thick solid, dashed and dash-dotted
lines stand for $(c^2_\alpha, c^2_\beta, c^2_\gamma)=(1,0,0)$,
$(0,1,0)$ and $(0,0,1)$, respectively.
Obviously, with proper values for the parameters, the decaying dark
matter could fit the Fermi-LAT and PAMELA data well simultaneously.

Inspired by the recent data measured by Fermi-LAT and PAMELA, we
have found that the puzzles of small neutrino masses and baryon
asymmetry as well as dark matter can be solved uniformly by
extending the lepton sector of the SM with two $Z_2$ symmetries. In
the proposed model, the conventional type-I and radiative
corrections seesaw mechanisms coexist to generate the masses of
neutrinos. The matter-antimatter asymmetry originated from
leptogenesis could be accomplished by the introduced new stuff
associated with soft $Z_2$ breaking terms. With the same soft
breaking effects, the condition of small couplings for a long-lived
decaying dark matter can be realized naturally through radiative
corrections.

\section*{Acknowledgements}
We would like to think Prof. Chao-Qiang Geng and Dr. Dmitry V.
Zhuridov for useful discussions. This work is supported in part by
the National Science Council of R.O.C. under Grant No:
NSC-97-2112-M-006-001-MY3.


\end{document}